\documentclass [journal] {IEEEtran}

\ifCLASSINFOpdf
\else
   \usepackage[dvips]{graphicx}
\fi
\usepackage{url}

\usepackage{graphicx}
\usepackage{amsmath}
\usepackage{array}

\begin{document}

\title{A Reversed and Shift Sparse Array Scheme Based on the Difference and Sum Co-array}

\author{Yan Zhou
\thanks{This work was sponsored in part by National Natural Science Foundation of China under Grants 61901371, 61971349 and 61801363, by the Natural Science Basic Research Program of Shaanxi under grant 2020JQ-600. (Corresponding author: Yan Zhou). }
\thanks{Yan Zhou is with the School of Information Science and Technology, Northwest University, Xi'an 710127, China (email: yanzhou@nwu.edu.cn)}}

\maketitle

\begin{abstract}
The reversed and shift (RAS) sparse array scheme, which is based on the difference and sum co-array (DSCA) and remarkably enhances the capability of identifying sources, is proposed. For the original nested array (NA) or co-prime array (CPA), there exists a large overlap between its difference co-array and sum co-array, which prevents it from obtaining high degrees of freedom (DOFs). Motivated by this fact, the RAS scheme is designed for reducing the overlap while increasing the DOFs for direction of arrival (DOA) estimation. The proposed scheme is effective for both NA and CPA. The closed-form expression for the relationship between the number of physical sensors and the number of consecutive DSCA sensors is derived. Compared with some representative DSCA based sparse arrays, the proposed one can achieve longer consecutive virtual array. Simulation experiments are carried out to exhibit the enhanced DOA estimation performance of RAS scheme.
\end{abstract}


\IEEEpeerreviewmaketitle

\section{Introduction}

\IEEEPARstart{I}{n} radar system, wireless communication, speech processing and so on, direction of arrival (DOA) estimation has been always receiving significant attention [1]-[5]. According to the parameter identification theory, by uniform linear array (ULA) with $T $ sensors, less than  $T$ sources can be resolved by the classical DOA estimation techniques. Recently, the co-array, which consists of virtual sensors generated by the sparse array with physical sensors, has attracted a great number of related researchers and practitioners due to its ability of resolving more sources than sensors. Based on this concept, plenty of difference co-array (DCA) based sparse arrays, including nested array (NA) [6] and co-prime array (CPA) [7], have been developed [8]-[14]. By combining the sum co-array (SCA) with DCA, the degrees of freedom (DOFs) available for DOA estimation can be further improved [15]-[20]. In [17], by virtual of adopting the vectorized conjugate augmented MUSIC (VCAM) algorithm, the difference and sum co-array(DSCA), which consists of DCA, non-negative SCA and non-positive SCA, is utilized to increase the DOFs.

In this paper, based on the concept of DSCA, a reversed and shift (RAS) sparse array scheme, which is effective for both NA and CPA, is designed for further increasing DOFs. We start from RAS scheme for NA (RAS-NA). For the original NA, there is a large overlap between its DCA and SCA. Based on this fact, NA is designed to be reversed and then shifted to a position for reducing the overlap while increasing the DOFs. Here, it is found that the most appropriate shifted distance is equal to the largest spacing in NA. The same scheme can also be applied into CPA and the RAS scheme for CPA (RAS-CPA) is configured as well. By comparing RAS-NA and RAS-CPA with other representative sparse arrays, such as NADiS [19] and DsCAMpS [20], the proposed scheme can achieve higher DOFs, which means that more resolved sources and higher DOA estimation accuracy can be realized. Finally, the numerical experiments, mainly with regard to the ability for resolving sources and the estimation accuracy, are carried out to verify the performance of RAS scheme.

\textbf{Notation:} We denote superscript $[ \bullet ]^T$ as transpose. $\max ( \bullet {\rm{)}}$ means the maximum value in a set. For any two sets of sensor positions ${\cal P}$ and ${\cal Q}$, the set operations with regard to them are listed here for a more appropriate understanding of the paper
\begin{equation}
\begin{aligned}
&Sum{\rm{ }}\ set{\rm{: }}{\cal P} + {\cal Q} = \{ p + q|\forall p \in {\cal P},{\rm{ }}q \in {\cal Q}\}, \\ 
&Difference{\rm{ }}\ set:{\rm{ }}{\cal P} - {\cal Q} = \{ p - q|\forall p \in {\cal P},{\rm{ }}q \in {\cal Q}\}, \\ 
&Translation{\rm{ }}\ Set:{\rm{ }}c + {\cal P} = \{ c + p|\forall p \in {\cal P},{\rm{ }}c \in {\cal Z}\}
\end{aligned}
\end{equation}
where $c $ is a constant and ${\cal Z}$ represents the integer set. The set $\{n:m\}$ represents a consecutive set ranging from $n$ to $m$.

\section{Signal Model}

The sensor positions of a $T$ sensor array subject to the following set
\begin{equation}
{\cal P} = \left\{ {p_1 d,p_2 d, \cdots ,p_t d \cdots ,p_T d,t = {\rm{1}},{\rm{2}}, \cdots ,T} \right\}
\end{equation}
where $p_t d$ represents the position of a sensor, $d = \lambda /2$ with $ \lambda$ denoting wavelength. The array output for $L$ far-field narrowband sources is
\begin{equation}
{\bf{x}}(t) = \sum\limits_{l = 1}^L {{\bf{a}}(\theta _l )s_l (t)}  + {\bf{\varepsilon }}(t) = {\bf{As}}(t) + {\bf{\varepsilon }}(t)
\end{equation}
in which ${\bf{s}}(t) = [s_1 (t), \cdots ,s_L (t)]^T $ contains the source signals,${\bf{\varepsilon }}(t)$ represents the Gaussian noise, ${\bf{a}}(\theta _l ) = [{\mathop{\rm e}\nolimits} ^{ - j2\pi p_1 d\sin (\theta _l )/\lambda )} , \cdots ,{\mathop{\rm e}\nolimits} ^{ - j2\pi p_T d\sin (\theta _l )/\lambda } ]^T $ is the steering vector with $\theta _l $, $l = 1,2, \cdots ,L$ being the incident angle and ${\bf{A}} = [{\bf{a}}(\theta _1 ), \cdots ,{\bf{a}}(\theta _L )]$ [1]. The sources are mutually uncorrelated. By utilizing the VCAM algorithm, a new virtual snapshot ${\bf{z}}$, whose elements correspond to the signal received at the DSCA of ${\cal P}$, can be extracted from the covariance matrix of ${\bf{x}}(t)$. The detailed introduction to VCAM can be seen in [17]. For a sparse array, its DSCA is defined as [15]
\begin{equation}
{\cal D} = \{ {\cal P} + {\cal P}\}  \cup  - \{ {\cal P} + {\cal P}\}  \cup \{ {\cal P} - {\cal P}\} 
\end{equation}
As the expression in (4), DSCA includes DCA, non-negative SCA and non-positive SCA. In VCAM, the consecutive segment in DSCA determines the virtual ULA or the DOFs available for DOA estimation. In the following, a new sparse array scheme that enhances the DOA estimation performance by using DSCA is proposed.  For convenience, $d$ is omitted in equations.

\section{The Proposed RAS Scheme}
\subsection{The design of RAS-NA}
 In general, with the total of $T = N + M$ sensors, the sensor positions in a NA subject to ${\cal P}_{NA} {\rm{ = }}{\cal P}_1  \cup {\cal P}_2 $ [7], where 
\begin{equation}
\begin{split}
&{\cal P}_1  = \left\{ {{\rm{0:}}N - 1} \right\}{\rm{ }}, \\
&{\cal P}_2  = \left\{ {N,2(N + 1) - 1, \cdots ,M(N + 1) - 1} \right\}
\end{split}
\end{equation}
As commonly known, the DCA of a NA is hole-free and ranging from $-(MN + M - 1)$ to $(MN + M - 1)$.Though the SCA of a NA is not hole-free, it still poetesses a consecutive segment ranging from $-(M+1)(N+1)+2$ to $(M+1)(N+1)+2$. Due to a large overlap between DCA and SCA, the consecutive segment in the DSCA of a NA is not much longer than that in its DCA or SCA, which is not beneficial to the DOA estimation. Here, for further improving the DOFs, the RAS scheme will be created to fully exploit the relationship between the consecutive segments in DCA and SCA. The main motivation behind the proposed RAS scheme is that the SCA can be shifted according to the length of the consecutive segment in DCA to reduce the overlap. Based on the idea, let $\alpha  = M(N + 1) - 1{\rm{ = }}\max ({\cal P}_{NA} {\rm{)}}$ be the shift factor, then the shift NA (S-NA) is deployed as ${\cal P}_{SN} {\rm{ = }}{\cal P}'_1 {\rm{ }} \cup {\cal P}_2 ^\prime   \cup {\cal P}_3 ^\prime$, where 
\begin{equation}
\begin{split}
{\cal P}_1 ^\prime   =& \{ 0\} {\rm{ }},\ {\cal P}'_2  = \left\{ {{\rm{0:}}N - 1} \right\} + (\alpha  + 1)/2, \\ 
 {\cal P}'_3  = &\left\{ {N,2(N + 1) - 1, \cdots ,M(N + 1) - 1} \right\} \\
 &+ (\alpha  + 1)/2 
\end{split}
\end{equation}
Compared with the original NA, the DCA of S-NA maintains the original position since the shift factor is inactive for the DCA. The SCA of S-NA is shifted to the end of DCA to increase the DOFs. Unfortunately, since the reference sensor in the original NA is shifted, a reference sensor must be added in S-NA. The new added sensor increases the number of sensors used in S-NA and hence S-NA fails to considerably enhance the DOFs available for DOA estimation. In order to avoid increasing the sensor and reduce the overlap, NA is reversed and then shifted to further optimize S-NA as RAS-NA, namely ${\cal P}_{RN} {\rm{ = }}{\cal P}''_1 {\rm{ }} \cup {\cal P}''$, where 
\begin{equation}
{\cal P}''_1  = \alpha  - {\cal P}_1 ,\  {\cal P}''_2  = \alpha  - {\cal P}_2 
\end{equation}
Based on this structure, the number of sensors maintains the same as in NA while the SCA of RAS-NA is shifted to reduce the overlap. The DCA of RAS-NA is still consecutive and ranging from $-(MN + M - 1)$ to $(MN + M - 1)$. By shifted, the consecutive parts in SCA are $\{  - 2M(N + 1) + 2:( - M + 1)(N + 1)\}$ and $\{ (M - 1)(N + 1):2M(N + 1) - 2\}$.The number of consecutive DSCA sensors in RAS-NA is $4MN + 4M - 3$, which is considerably larger than that in NA. For clearly showing the improvement, RAS-NA is compared with NA and NADiS [19] in Fig. 1. Among the sparse arrays considered here, RAS-NA can always outperform its counterparts.
\begin{figure}
\centerline{\includegraphics[width=\columnwidth]{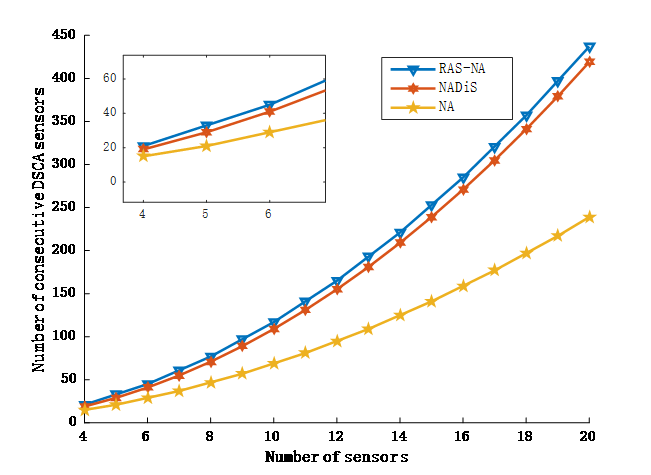}}
\caption{The number of consecutive DSCA sensors comparison for NA based sparse array}
\end{figure}

\subsection{The design of RAS-CPA}
As the definition, with the total of $T = 2N + M - {\rm{1}}$ sensors, the array deployment in CPA is ${\cal Q}_{CA} {\rm{ = }}{\cal Q}_1  \cup {\cal Q}_2 $, where
\begin{equation}
{\cal Q}_1  = \left\{ {0:M - 1} \right\}N,\ {\rm{ }}{\cal Q}_2  = \left\{ {0:2N - 1} \right\}M
\end{equation}
$N$ and $M$ are co-prime numbers satisfying $N<M$ [9]. It is well-known that the DCA of a CPA can generate a consecutive virtual ULA ranging from $- MN - N + 1$ to $ MN + N - 1$. Though the SCA of a CPA is well illustrated in [16]-[17], it is not sufficiently exploited to improve the DOFs. Here, inspired by the design of RAS-NA, the RAS-CPA is arranged as
\begin{equation}
{\cal Q}_{RC} {\rm{ = }}\beta  - {\cal Q}_{CA} 
\end{equation}
where $\beta  = \max ({\cal Q}_{CA} ) = (2N - 1)M$ is the shift factor for RAS-CPA. For demonstrating that the proposed scheme will remarkably outperform CPA in the number of consecutive DSCA sensors, the \emph{Proposition 1} is introduced first.

\emph{Proposition 1}: The SCA of a CPA, which is generated  according to co-prime numbers $N$ and $M$ and $M-N=1$, contains a consecutive virtual array ranging from $(N-1)N$ to $(2N - 1)(M + 1) + 1$.

\emph{Proof}:All elements in ${\cal Q}_1  + {\cal Q}_2$ can be listed in a matrix as
\begin{equation}
\begin{scriptsize}
\left[ {\begin{array}{*{4}c}
   {(2N - 1)M} &  \cdots  & M & 0  \\
   {N + (2N - 1)M} &  \cdots  & {N + M} & N  \\
    \vdots  &  \cdots  &  \vdots  &  \vdots   \\
   {(M - 1)N + (2N - 1)M} &  \cdots  & {(M - 1)N + M} & {(M - 1)N}  \\
\end{array}} \right]
\end{scriptsize}
\end{equation}
It should be noted that the elements along every diagonal can be expressed in the set ${\cal M} = \bigcup\limits_{i = 1}^{2N - 1} {{\cal M}_1 (i)} \bigcup\limits_{i' = 1}^{M - 1} {{\cal M}_2 (i')}$, where
\begin{equation}
\begin{split}
&{\cal M}_1 (i) = \\
&\left\{ {(i - k)N + kM,{\mathop{\rm sgn}} (i - M + 1) \le k \le i,k \in {\cal Z}} \right\}
\end{split}
\end{equation}
\begin{equation}
\begin{split}
&{\cal M}_2 (i') = \\
&\left\{ {k'N + (2N - 1 + i' - k')M,i' \le k' \le M - 1,k' \in {\cal Z}} \right\},
\end{split}
\end{equation}
in which the function 
${\mathop{\rm sgn}} (x) = \left\{ \begin{array}{l}
 x,{\rm{ if }}x \ge {\rm{0}} \\ 
 0,{\rm{ else}} \\ 
 \end{array} \right.
$. Obviously, ${\cal M}_1 (i)$ and ${\cal M}_2 (i')$ are both consecutive sets since the difference of adjacent elements in each set is
\begin{equation}
\begin{split}
&(i - k)N + kM - (i - k + 1)N + (k - 1)M\\
& = M - N = 1
\end{split}
\end{equation}
Based on this fact, as long as $[i + 1 - {\mathop{\rm sgn}} (i - M + 2)]N + {\mathop{\rm sgn}} (i - M + 2)M - iM \le 1$, i. e.
$i \ge N - 1
$, $
\bigcup\limits_{i = 1}^{2N - 1} {{\cal M}_1 (i)} 
$
 contains a consecutive set $
\left\{ {(N - 1)N:(2N - 1)M} \right\}
$
.In addition, it is noted that by combining $
{\cal M}_2 (1)
$
and $
{\cal M}_2 (2)
$, namely $
\{ (M - 1)N + (2N - M + 1)M, \cdots ,N + (2N - 1)M\} 
$
 and $
\{ (M - 1)N + (2N - M + 2)M, \cdots ,2N + (2N - 1)M\} 
$
,with $
\left\{ {(N - 1)N:(2N - 1)M} \right\}
$
 and the element $\{2NM\}$ in $
{\cal Q}_1 {\rm{ + }}{\cal Q}_1 
$
 together, one can find that they actually form a consecutive set $
\left\{ {(N - 1)N:2N + (2N - 1)M} \right\}
$. Therefore, the SCA of the CPA with $M-N=1$ contains a consecutive segment from $(N-1)N$ to $2N+(2N-1)M$, and the proof is completed. Next, the DSCA of RAS-CPA will be illustrated.

\emph{Proposition 2}: The DSCA of RAS-CPA with $M-N=1$, contains a consecutive virtual array ranging from $-4MN+2M+N(N-1)$ to $4MN-2M-N(N-1)$.

\emph{Proof}: As the above analysis, for RAS-CPA, the union of its DCA and SCA is not necessarily a consecutive set since $2MN + 2N - M > MN + N - 1$ when $N > 4$. For completing the proof, some short consecutive sets exist in DCA and SCA are required. As the beginning, the items in the difference set of two subsets, namely ${\rm{\{ }}N + 1:2N - 1{\rm{\} }}M$ and ${\rm{\{ }}0:M - 1{\rm{\} }}N$, in CPA is listed in a matrix as
\begin{equation}
\begin{tiny}
NM + \left[ {\begin{array}{*{20}c}
   M & {M - N} &  \cdots  & {M - (M - 1)N}  \\
   {2M} & {2M - N} &  \cdots  & {2M - (M - 1)N}  \\
    \vdots  &  \vdots  &  \cdots  &  \vdots   \\
   {(N - 1)M} & {(N - 1)M - N} &  \cdots  & {(N - 1)M - (M - 1)N}  \\
\end{array}} \right]
\end{tiny}
\end{equation}
Note that $M-N=1$ and hence the items along the diagonals located at the lower triangle area form the following short consecutive sets
\begin{equation}
\begin{split}
 &{\cal M}_3 (i'') = \\
&\left\{ {NM + i''M + N - 1 - k'',i'' \le k'' \le N - 1,k'' \in {\cal Z}} \right\}, \\ 
 &1 \le i'' \le N - 1,i'' \in {\cal Z} \\ 
\end{split}
\end{equation}
Now, recall the short consecutive sets in (12) and the shifted non-negative SCA subsets corresponding to (12) can be rearranged as 
\begin{equation}
\begin{split}
 &{\cal M}_4 (i') = \\
&\left\{ {NM + (i' - 1)M + k',N - i' \le k' \le N,k' \in {\cal Z}} \right\}, \\ 
 &1 \le i' \le N - 1,i' \in {\cal Z} \\ 
\end{split}
\end{equation}
Let $i'' = i'$, the union of ${\cal M}_3 (i')$ and ${\cal M}_4 (i')$ is expressed as
\begin{equation}
\begin{aligned}
\begin{array}{l}
 {\cal M}_4 (i' - 1) \cup {\cal M}_4 (i') \cup {\cal M}_3 (i' - 1) \cup {\cal M}_3 (i') = \\ 
 \ \left\{ {NM + (i' - 2)M + N - i' + 1:NM + i'M + N - i'} 
\right. 
\\  
\left.  \ \, {- 1} \right\}. \\  
2 \le i' \le N - 1,i' \in {\cal Z}  \\ 
 \end{array}
\end{aligned}
\end{equation}
Evidently, the set in (17) ranging from $NM+N-1$ to $2NM-M$ can fill the gap between the  DCA of RAS-CPA and the non-negative SCA of it. Due to the symmetry of DCSA, a consecutive set ranging from $-2NM+M$ to $-NM-N+1$ can also be found in the DCSA of RAS-CPA, which can fill the gap between the DCA of RAS-CPA and the non-positive SCA of it. As a consequence, the proof is completed.
\begin{figure}
\centerline{\includegraphics[width=\columnwidth]{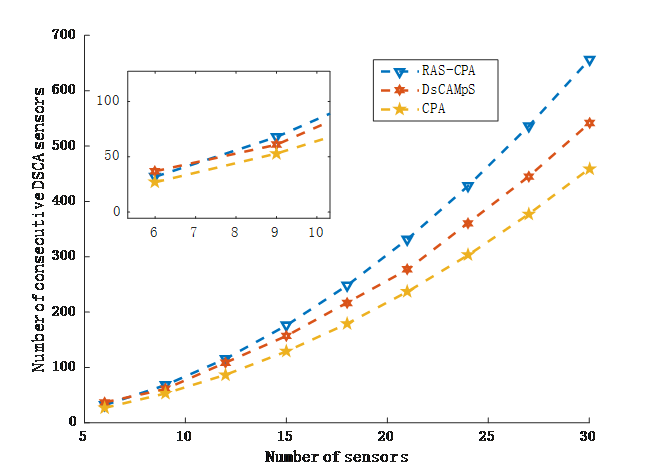}}
\caption{The number of consecutive DSCA sensors comparison for CPA based sparse array}
\end{figure}

With the total of $T=2N+M-{\rm{1}}$ sensors, the number of consecutive DSCA sensors in RAS-CPA with $M-N={\rm{1}}$ can reach as many as ${\rm{8}}MN - {\rm{4}}M - {\rm{2}}N(N - 1){\rm{ + 1}}$. For clearly exhibiting the advantage of RAS-CPA, it is compared with CPA and DsCAMpS [20] in Fig. 2. Except when $T={\rm{6}}$ RAS-CPA can always outperform its counterparts.

\section{Simulation Results}
 In the simulations, $Q$ sources are uniformly located from $- 50^ \circ  $ to $50^ \circ  $. The VCAM algorithm is employed here to resolve sources [17]. The number of consecutive DSCA sensors in NA, NADiS, RAS-NA, CPA, DsCAMpS and RAS-CPA is denoted as $u_{NA}$, $u_{ND}$, $u_{RN}$, $u_{CA}$,$u_{DC}$ and $u_{RC}$, respectively.

\subsection{The NA based sparse array comparison}
 In this subsection, $8$ physical sensors are deployed.
Accordingly, $u_{NA}=47$, $u_{ND}=71$ and $u_{RN}=77$. The MUSIC spectrum is shown in Fig. 3. There are $26$ sources with $0dB$ SNR (signal to noise ratio) and the red dot lines exhibit the true incident angle of each source. The number of snapshots is $300$. It is clearly shown that RAS-NA correctly resolves all the $26$ sources while NADiS fails to achieve it. In addition, the MUSIC spectrum of NA cannot be plotted here due to the insufficient noise subspace dimension.
\begin{figure}
\centerline{\includegraphics[width=\columnwidth]{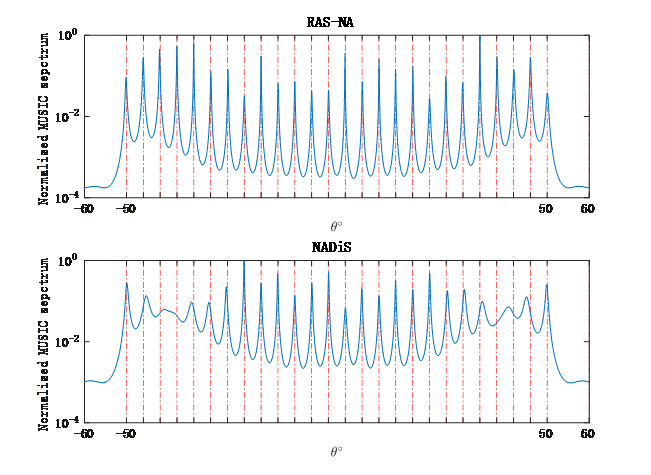}}
\caption{The MUSIC spectrum comparison for NA based sparse array}
\end{figure}
\begin{figure}
\centerline{\includegraphics[width=\columnwidth]{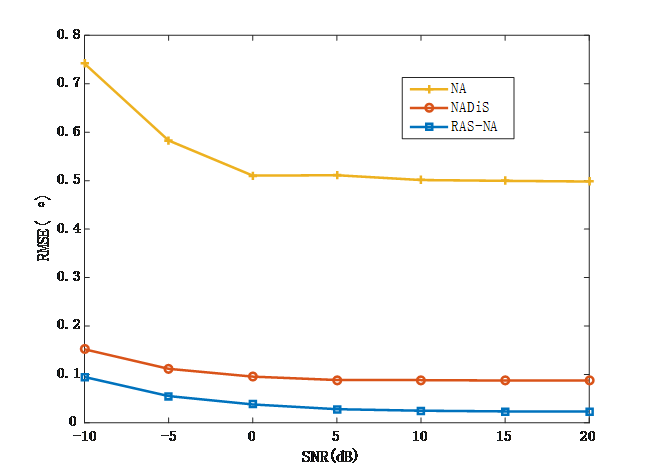}}
\caption{RMSE varies with SNR for NA based sparse array}
\end{figure}

For evaluating the estimation accuracy of each sparse array, the DOA estimation root mean-square error $RMSE = \sqrt {\sum\limits_{l = 1}^L {(\hat \theta _l  - \theta _l )^2 /L} } 
$
, where $\hat \theta _l $ is the estimate of $\theta _l $, is averaged through 100 independent Monte Carlo runs. Here, 20 sources and 300 snapshots are considered. Apparently, the estimation accuracy improves with the increase of SNR in Fig. 4. For the full range of SNR, RAS-NA can always reach the lowest RMSE. 

\subsection{The CPA based sparse array comparison}
 In this subsection, $9$ physical sensors are exploited in each sparse array. Accordingly,$u_{CA}=53$ , $u_{DC}=61$ and $u_{RC}=69$. The MUSIC spectrum is demonstrated in Fig. 5, where $27$ uncorrelated sources with $0dB$ SNR are considered and the number of snapshots is $300$. Under this condition, RAS-CPA has succeeded in resolving the $27$ sources while DsCAMpS have failed to resolve all the sources. Finally, for evaluating the estimation accuracy of the CPA based sparse array, the DOA estimation RMSE is obtained through $100$ independent Monte Carlo runs. Here, $20$ sources are considered. As shown in Fig. 6, during the full range of SNR, the proposed RAS-CPA can always reach the lowest RMSE.
\begin{figure}
\centerline{\includegraphics[width=\columnwidth]{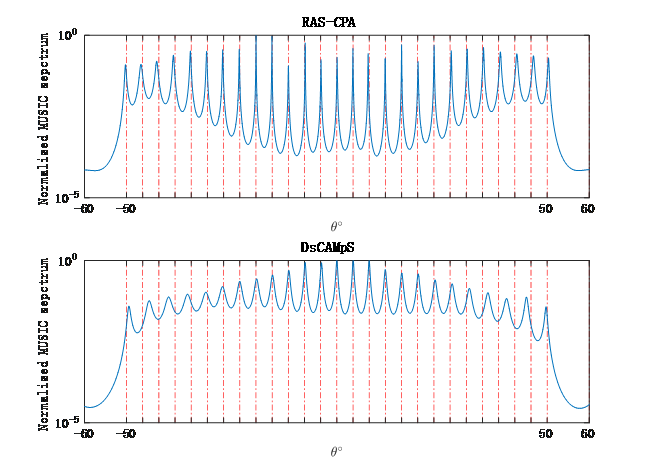}}
\caption{The MUSIC spectrum comparison for CPA based sparse array}
\end{figure}
\begin{figure}
\centerline{\includegraphics[width=\columnwidth]{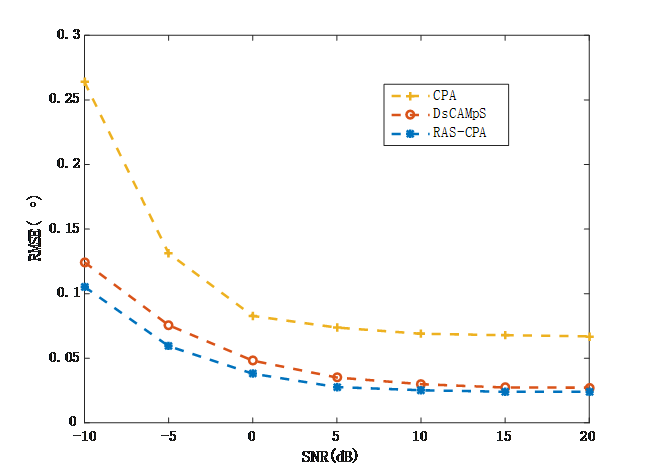}}
\caption{RMSE varies with SNR for CPA based sparse array}
\end{figure}

\section{Conclusions }
 A new sparse array scheme, which is based on the concept of DSCA, has been proposed for increasing the consecutive DSCA elements in NA and CPA. By properly utilizing the relationship between the DCA and SCA, the original NA is reversed and then shifted to a position such that more consecutive DSCA elements can be achieved. The same structure has been found to be effective for CPA. Given the same number of physical sensors, RAS scheme possesses larger number of consecutive DSCA elements, hence higher DOFs available for DOA estimation than its counterparts. Simulation results verify the performance of the proposed RAS scheme. In the future work, we will try to embed other explicitly expressed sparse arrays into RAS scheme.

\end{document}